# The Global Impact of AI-Artificial Intelligence: Recent Advances and Future Directions, A Review

Chandregowda (Chandra), Pachegowda
Department of Professional Security Studies
New Jersey City University
Jersey City, New Jersey, USA
ORCID: https://orcid.org/0009-0006-5841-6377

*Abstract*— Artificial intelligence (AI) is an emerging technology that has the potential to transform many aspects of society, including the economy, healthcare, and transportation. This article synthesizes recent research literature on the global impact of AI, exploring its potential benefits and risks. The article highlights the implications of AI, including its impact on economic, ethical, social, security & privacy, and job displacement aspects. It discusses the ethical concerns surrounding AI development, including issues of bias, security, and privacy violations. To ensure the responsible development and deployment of AI, collaboration between government, industry, and academia is essential. The article concludes by emphasizing the importance of public engagement and education to promote awareness and understanding of AI's impact on society at large.

## I. Introduction

Artificial intelligence (AI) is a rapidly evolving technology in recent years. It has the potential to transform the world in several aspects of society, across all sectors of industries and governments. AI has become the most spoken and crucial topic of interest in research and development in the academic field, mainly around the technological and natural language processing space.

Figure-1 illustrates the AI-related publications from the year 2000-2022, it shows the attention from the corner that it is getting.

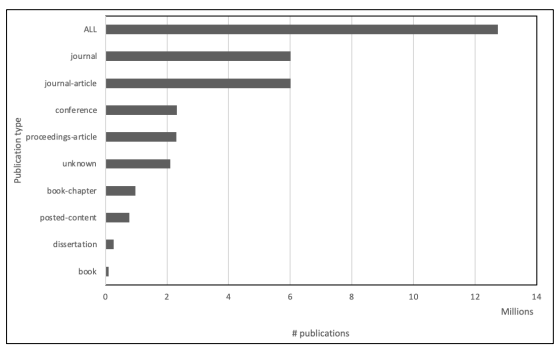

Figure-1. Data source: OpenAlex [1]

According to Russell S. J. and Norvig P. [2], Artificial intelligence (AI) is a field of computer science focused mainly on creating a machine or a program that performs the tasks that typically require human intelligence. For example, machine learning, problem-solving, decision-making, understanding languages, natural language processing, automated reasoning, and carrying out tasks equal to humans. These AI attributes have been broadly classified into four approaches.

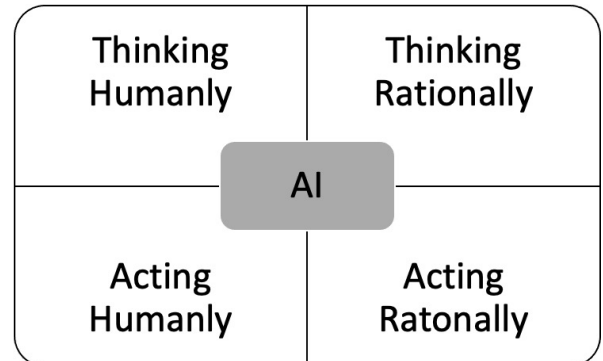

Figure-2. Table information source: [2]

- Thinking humanly: is a cognitive model approach where a machine is trained to think like a human and respond. This supports Haugeland J.'s [3] definition which states there are efforts to make the computer think like humans, 'machines with minds' to be accurate. Automated activities like decision-making, problem-solving, and learning are associated with human thinking [4].
- Think rationally: is a logical thinking approach using laws of thought - right thinking that is irrefutable. This is nothing but sensible thinking through the use of computational models and patterns [5]. The computations help perceive, reason, and act [6].
- Acting humanly: the computer is expected to possess capabilities such as natural language processing (NLP), knowledge representation (store learnings), automated reasoning (answering based-on information stored), and machine learning (adapt, detect, and infer). Machines have the art of performing functions that require human intelligence [7]. Machines do things that people are better at [8].
- Act rationally: leverage rational thinking through correct inferences drawn from laws of thought. Intelligent agents act using computational intelligence [9]. Acting with adapting intelligence behavior [10].

In today's world, AI has been used in numerous applications such as medical diagnosis, autonomous cars, home automation, social media, chatbots, virtual assistance, financial analysis, and there are many more. According to the world economic forum paper - The New Physics of Financial Services - 'understanding how Artificial Intelligence is Transforming the financial ecosystem'[11], people talk about AI in the context of capabilities that can allow them to run businesses in a new way with certain predictable power with a degree of autonomous learnings to accelerate tasks across all work-area rather a sheer technological approach or a well-defined computer science process framework. In addition, the report highlights that AI does not exist in a vacuum, it is intertwined with all the technological advances. [12] The best example is that AI has been started with industrial robotics to carry out a specific set of tasks based on pre-loaded instructions and gained momentum with home automation applications such as conversational or voice-enabled devices, appliances, sensors, and robotic cameras. AI's advancement has strongly convinced scholars to provide benefits in multi-folds and it is being used across all sectors. Simply put, in a broader context AI is going to be here to disrupt the world in all fields, help humans provide a better experience, improved productivity, and accelerate innovation.

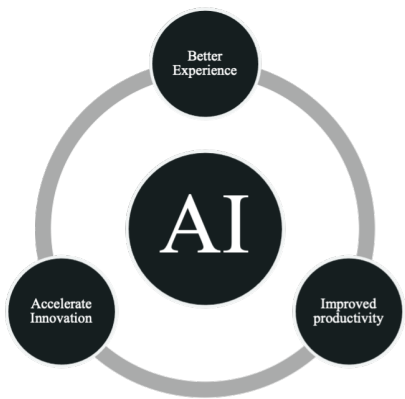

Figure-3

Although AI had made its ground to prove benefits in many fields, it can also pose several challenges and ethical concerns if the technology has been misused. This paper aims to explore and synthesize the global impact of AI in recent years by reviewing literature and discussing the impacts of AI on various aspects of human lives.

## II. IMPLICATIONS OF AI

### A. Economic implications

Recent studies and developments show, AI has the potential to increase economic growth through improved labor productivity, reduced resource consumption, accelerate innovations across sectors and touch human lives in many aspects for the better good. It has the potential to revolutionize the global economy, healthcare, and society at large. In the past few years: AI has gained significant attention in the academic and industry domains, leading to several studies on its impact. [13] Microsoft's recent investment of billions of dollars in the OpenAI platform products (ChatGPT, Co-Pilot) and [14] Google's Bard - The next generation of AI for developers, release for public experiments are the best examples of its predictable economic growth and development in the software industry sector. AI has both positive and negative effects on the economy. AI can improve efficiency and productivity, leading to increased economic growth and job creation. On the other hand, AI can also lead to job displacement and increase the risk of economic inequality. According to a review of the economics of artificial intelligence by Lu Y. and Zhou Y. [15], AI certain extent would rescind jobs and increase inequality.

AI has shown promise in the field of healthcare. The studies have explored the impact of AI on healthcare and found that AI can lead to improved accuracy in diagnosis, enhanced patient outcomes, reduced medical errors, better resource utilization, and increased efficiency in healthcare delivery. Additionally, AI can lead to cost savings and increased accessibility to healthcare services. Another industry where AI has made noteworthy strides is finance. According to a 2018 report, "The New Physics of Financial Services" by the World Economic Forum [11], AI has the potential to transform the financial sector by enhancing customer service, improving risk management, and reducing fraud. AI can also improve the accuracy of financial forecasts and decision-making, leading to better investment outcomes.

McKinsey's global survey on AI shows more than a hundred percentile growth.

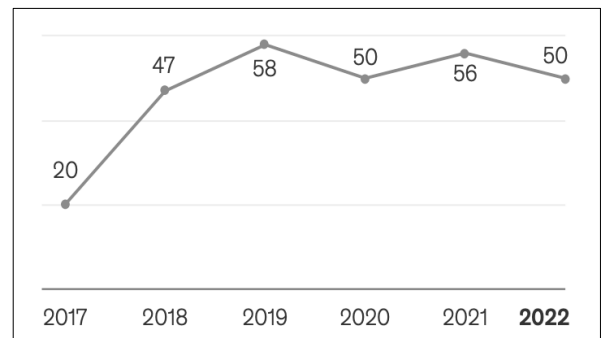

Figure-4. Chart Credit: [16]

AI adoption has more than doubled since 2017 and the proportion of organizations using AI is between 50 and 60 percent over the past few years [16].

The Whitehouse has recently released an economic study in response to the US-EU Trade and Technology Council Inaugural Joint Statement, which examines the impact of artificial intelligence on the future of the workforce in the European Union and the United States of America [17]. The study reveals that AI adoption is evident in firms regardless of their size and across industries such as finance, information, professional services, and management, and more importantly, the report emphasizes that younger and large firms are more likely to adopt AI [17].

However, AI's impact on the physics of financial services will demand increased collaboration to address the emerging uncertainties as its utilization grows [11]. Whereas in security, AI globally creates new types of risks in financial services systems and new risk-management and mitigation strategies that will be required. In the area of human capital, AI creates new forms of human labor needs and displaces portions of the labor force. Collective action by institutions and regulators will be required. As AI automates decision-making processes, new methods of protecting consumers and ensuring the public interest is safeguarded will be required for consumer protection. Most importantly, industry leaders, regulators, and public policy institutions must explore and address emerging social issues [11]. As Scherer [18] points out, regulating AI systems poses several challenges, including determining which competencies and strategies are necessary to mitigate risks associated with AI. Besides, studies found that ethical considerations are essential in AI and automation, given the potential for bias and discrimination in decision-making processes.

*B. Ethical Implications*

There are growing concerns about AI's ethical implications becoming increasingly integrated into our lives. The ethical implications of AI range from issues surrounding bias, privacy & security, and accountability to transparency, explainability, and human rights. [19], in their study 'The ethics of AI', argue that the ethical implications of AI require careful consideration of its ability to affect human behavior and values. The use of AI in social media platforms can lead to the spread of misinformation and fake news, undermining democratic processes and social cohesion. To mitigate these issues, researchers have proposed using crowdsourcing and other approaches to combat misinformation on social media [20]. Another ethical consideration is the potential for AI to reinforce and amplify existing biases in society and amplify false information as truth. The design of AI systems must consider the potential for biased outcomes and work toward developing transparent and accountable AI systems. There is an increasing need for greater transparency and interdisciplinary collaboration to address the ethical, legal, and social implications and explainability of AI systems so that their decision-making processes can be understood and scrutinized by humans. The interdisciplinary approach is also necessary to address the complex ethical considerations associated with AI, including, issues related to privacy, bias, and transparency. Further, an empathetic approach to AI could address these concerns, leading to more ethical AI systems.

*C. Social Implications*

The social implications of artificial intelligence are huge and wide-ranging, with applications in education, media, and transportation such as autonomous vehicles and more. In education, AI can be used to personalize learning and improve educational outcomes for students. The policymakers in public and private sectors should consider investing in AI education and training programs to ensure that the workforce is equipped with the necessary skills to work alongside AI systems. This investment will also help to mitigate concerns related to job displacement due to automation.

Despite the benefits, AI also presents several challenges, including the potential for biased and misinformation on social media platforms [20] and the need for ethical, legal, and technical governance of AI systems [21]. On the internet and social media space, AI is used for content recommendation and personalized advertising, leading to concerns about the potential for manipulation and the creation of filter bubbles - 'A situation in which an internet user encounters only information and opinions that conform to and reinforce their own beliefs, caused by algorithms that personalize an individual's online experience' [22]. In addition, the use of AI in these media platforms can also raise privacy concerns as it can collect vast amounts of personal data without consent. Therefore, it is crucial to consider the potential positive and negative impacts of AI and to work towards responsible development and implementation of AI systems. In the transportation space, adopting AI-based technologies has already shown great potential to improve safety, efficiency, accessibility, and maintainability. Self-driving cars, for instance, have the potential to significantly reduce accidents and fatalities caused by human errors. Conversely, there are still concerns about the safety of autonomous (self-driving) vehicles and the need for greater regulation and supervision.

*D. Security and Privacy Implications*

A significant challenge with AI is the potential for security and privacy breaches. As AI systems become more prevalent in various industries, they may become prime targets for cyberattacks, which can compromise sensitive data and information. In addition, since AI has already been integrated into our daily lives, there is a need for clear and comprehensive regulation. For example, in healthcare, AI systems that store patient data could be vulnerable to cyberattacks, leading to the theft of patient information. Similarly, AI systems used in finance may be vulnerable to fraud, resulting in financial losses for individuals and organizations. On the other hand, bad attackers can use AI for sophisticated and powerful cybersecurity attacks. To mitigate security and privacy concerns, it is essential to implement robust security measures, such as encryption and multi-factor authentication, to protect sensitive data. Regular security audits can also help identify and address vulnerabilities in AI systems.

The impact of AI on the global economy, healthcare, and society will continue to grow in the coming years. As such, it is crucial to consider the potential implications of AI carefully and work towards responsible development and implementation of AI systems. Scherer [18] highlights the importance of developing regulatory frameworks that balance the benefits of AI with potential risks associated with its use, such as bias and discrimination. Additionally, AI developers and policymakers should consider security & privacy considerations from the outset of AI development to ensure that AI systems are designed with ethical principles in mind. The studies suggest the need for collaboration between healthcare providers and AI developers to ensure that AI systems are integrated seamlessly into existing healthcare systems while also ensuring patient safety and privacy. Further, policymakers should consider investing in AI education and training programs to ensure that the workforce is equipped with the necessary skills to work alongside AI systems. This investment could also help to mitigate concerns related to job displacement due to automation.

### III. CONCLUSION

In conclusion, the global impact of artificial intelligence is a multifaceted and complex topic that requires devotion from various stakeholders from both private and government institutions. AI has the potential to revolutionize many industries, including healthcare, finance, transportation, information, and content-sharing spaces such as the Internet, media, and social media. AI has huge potential to contribute to economic growth. To fully realize the benefits of AI and mitigate its risks, a collaboration between government, industry, and academia is essential. Regulatory frameworks should be established to ensure ethical and responsible AI development and international cooperation should be promoted to address global challenges posed by AI. Public engagement and education are also critical to promote awareness and understanding of AI's impact on society.

The future of AI is both exciting and uncertain, but careful consideration and proactive measures with appropriate regulation can help create a world where AI is harnessed for the greater good of humanity. Working collectively to ensure that AI is used for the benefit of all people rather than just a select few is vital. Through interdisciplinary collaboration and a focus on ethical considerations, AI has the potential to improve many aspects of our lives while also mitigating potential risks associated with its use.